\documentclass[preprint,12pt]{elsarticle}

\usepackage{amssymb}

 \usepackage{lineno}

\journal{JINST}

\biboptions{sort&compress}

\begin{document}

\begin{frontmatter}

\title{System upgrade for $\mu$Bq/m$^3$ level $^{222}$Rn concentration measurement}

\renewcommand{\thefootnote}{\fnsymbol{footnote}}
\author{
Y.~Liu$^{a}$, 
Y.P.~Zhang$^{b,c,d}$, 
J.C.~Liu$^{b,c,d}$, 
C.~Guo$^{b,c,d}\footnote{Corresponding author. Tel:~+86-1088236256. E-mail address: guocong@ihep.ac.cn (C.~Guo). }$, 
C.G.~Yang$^{b,c,d}$,
P.~Zhang$^{b,c,d}$,
Q.~Tang$^{a}\footnote{Corresponding author. Tel:~+86-1088236095. E-mail address: tangquan528@sina.com(Q.~Tang).}$,
Z.F.~Xu$^{a}$,
C.~Li$^{a}$, 
T.Y.~Guan$^{a}$,
S.B.~Wang$^{b}$
}

\address{
	${^a}${School of Nuclear Science and Technology, University of South China, Hengyang, China}

	${^b}${Experimental Physics Division, Institute of High Energy Physics, Chinese Academy of Sciences, Beijing, China}

	${^c}${School of Physics, University of Chinese Academy of Sciences, Beijing, China} 

	${^d}${State Key Laboratory of Particle Detection and Electronics, Beijing, China}

}


\begin{abstract}
	The Jiangmen Underground Neutrino Observatory (JUNO), a 20 kton multipurpose underground liquid scintillator detector, was proposed for the determination of the neutrino mass hierarchy as primary physics goal. The central detector will be submerged in a water Cherenkov detector to lower the background from the environment and cosmic muons. Radon is one of the primary background sources. Nitrogen will be used in several sub-systems, and a highly sensitive radon detector has to be developed to measure its radon concentration. A system has been developed based on $^{222}$Rn enrichment of activated carbon and $^{222}$Rn detection based on the electrostatic collection. This paper presents the detail of a $\mu$Bq/m$^3$ level $^{222}$Rn concentration measurement system and gives detailed information about how the adsorption coefficient was measured and how the temperature, flow rate, and $^{222}$Rn concentration affect the adsorption coefficient. 
\end{abstract}

\begin{keyword}
	Radon, Activated carbon, Adsorption coefficient, Low temperature 
\end{keyword}

\end{frontmatter}


\section{Introduction}\label{sec:section1}
The Jiangmen Underground Neutrino Observatory (JUNO) is a multipurpose neutrino experiment designed to determine the neutrino mass hierarchy and precisely measure the oscillation parameters by detecting reactor anti-neutrinos from the Yangjiang and Taishan Nuclear Power Plants with a 20 kton liquid scintillator (LS) detector located 700-m underground~\cite{JUNO_1, JUNO_2}. The natural radioactive gas radon ($^{222}$Rn), soluble in water and LS, is one of JUNO's most crucial background sources.

As an easily available low radioactive gas, nitrogen is used in several sub-systems of JUNO. For LS purification, nitrogen will be used as the stripping gas in the stripping plant to remove the radioactive gas dissolved in LS~\cite{LS}. For low radioactive ultrapure water production, nitrogen will be loaded into the water to improve the radon removal efficiency of degassing membrane~\cite{JUNO_2, JUNO_water}. During the detector operation, nitrogen will be filled between the water surface and the cover to prevent radon contamination from outside~\cite{JUNO_2}. The residual radon in nitrogen, which will introduce extra background to the detector, should be measured before use. Among these applications, LS purification has the highest requirement for nitrogen, in which the $^{222}$Rn concentration needs to be reduced to less than 30~$\mu$Bq/m$^3$. The radon concentration measurement system in this work is specially developed for measuring the radon concentration in nitrogen gas.

In Our previous work~\cite{Previous_1}, we measured the background of Saratech activated carbon and verified its radon adsorption performance. We investigated its optimal working conditions in this work and applied it to the highly sensitive radon concentration measurement. This paper is organized as follows. Section 2 describes the update of the measurement system. Section 3 describes how the adsorption coefficient of activated carbon is measured. Section 4 presents how the temperature, the gas flow rate, and the radon concentration in the gas affect the adsorption coefficient. Section 5 gives the $^{222}$Rn concentration measurement sensitivity of the system, section 6 presents discussions about the results and section 7 gives the conclusion and future prospects.

\section{Experimental principle and setup}\label{sec:section2}

In the highly sensitive $^{222}$Rn concentration measurement system, the authors first use low-temperature activated carbon to enrich the radon gas~\cite{AC_1, AC_2}, and then use a radon detector~\cite{SuperK_1, SuperK_2, SuperK_3, Borexino_2} to measure the radon concentration of the enriched gas. Activated carbon is an excellent adsorbent with a strong adsorption capacity for radon and has been widely used for radon removal in low background experiments~\cite{DS, SuperK, Xmass}. The adsorption capacity of activated carbon working at low temperatures to radon gas is significantly enhanced and it has been confirmed by many researchers~\cite{DS, Xmass, ACfiber, Borexino, Peking} and our previous work~\cite{Previous_1}. Based on the earlier results, the Saratech activated carbon~\cite{Saratech} is used in this work because of its excellent radon adsorption capability and low radioactive background.

\begin{figure}[htbp]
\centering
\includegraphics[width=12cm]{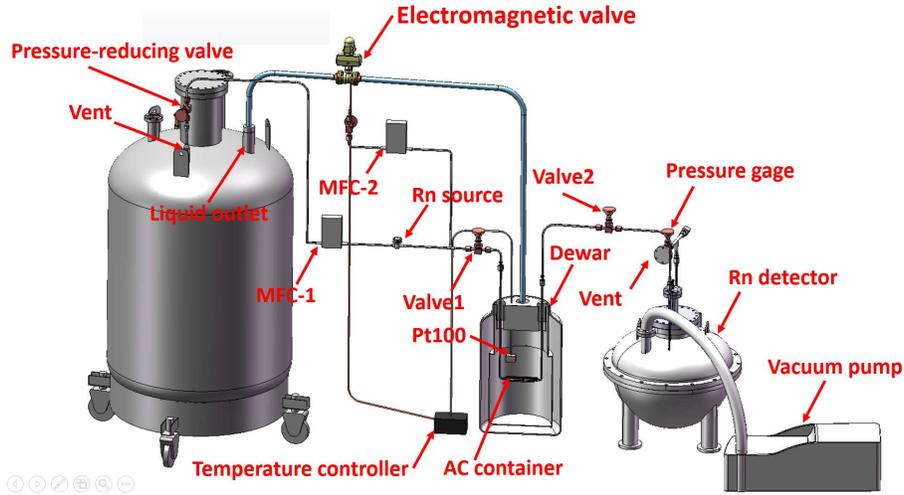}
\caption{\label{detector} Schematic system diagram, which consists of a high-pressure nitrogen tank, a gas flow control system, a gas flow radon source, an activated carbon container, a temperature control system, a vacuum pump, and a radon detector.}
\end{figure} 

Fig.~\ref{detector} shows the schematic of the system, which is updated from our previous work ~\cite{Previous_1}.  Based on the measurement requirements of this project, the details and the improvements are as follows:

(A) A high-pressure liquid nitrogen tank is used to increase the gas flow rate during the tests. The tank is filled with liquid nitrogen and its vent supplies evaporated nitrogen. The inside relative pressure is 12 bar. A pressure-reducing valve is connected to the vent, which can adjust the relative pressure of the outlet gas from 0 to 12 bar.  The liquid outlet of the tank is used to supply liquid nitrogen to the temperature control system.

(B) The gas flow control system consists of two mass flow controllers (MFC) and the relevant pipelines and valves. MFC-1 (1179A, MKS) is used to control the ﬂow rate of the gas through the gas flow radon source to keep the radon concentration in the gas stable and adjust the radon concentration in the gas. MFC-2 (1179A, MKS) is used to control the flow rate of the gas through the activated carbon to monitor the gas volume flow through the activated carbon.

(C) The gas flow radon source is made from BaRa(CO$_2$)$_3$ powder by the radon laboratory of South China University. The $^{222}$Rn concentration in the gas is inversely proportional to the flow rate. The tests used two $^{222}$Rn sources with different activities to obtain different $^{222}$Rn gas concentrations.

(D) The activated carbon container is a stainless steel pipeline with a length of 15~cm and a diameter of 1/4~inch. Two filter gaskets are placed at both ends. The aperture size of the gaskets is 60~$\mu$m. The container is placed in a dewar, and the temperature control system controls the temperature inside the dewar. The aperture size of the gaskets became much smaller than before because in previous tests we found that there was activated carbon debris entering the detector with the gas flow and contaminating the measurements.

(E)The temperature control system consists of a dewar, a temperature controller, a temperature sensor (Pt100), an electromagnetic valve, and a heating belt. The Pt100 temperature sensor placed next to the activated carbon chamber serves as the input to the temperature controller. The temperature controller controls the electromagnetic valve's switch to realize the intermittent injection of liquid nitrogen into the dewar. The liquid nitrogen vaporizes in the dewar, and the cooled nitrogen gas can cool down the activated carbon. The temperature controller also can control the switch of the heating belt, which is wound around the activated charcoal chamber. The heating belt is made of silicone rubber and heater strip and thus can withstand a low temperature of -196$^{\circ}$C and a high temperature of 200$^{\circ}$C. This system can realize a temperature change inside the dewar from -196$^{\circ}$C to 200$^{\circ}$C, and the temperature stability is within ±5 $^{\circ}$C. Different from our previous work~\cite{Previous_1}, this device needs to measure the radon concentration in the adsorbed gas, so a heating belt was used for the desorption of radon gas.

(F) The vacuum pump (ACP40, Pfeiffer ) is used to vacuum the detector before the radon gas is transferred from the activated carbon to the detector. With a pump, the residual radon gas inside the detector can be quickly removed. 

\begin{figure}[htbp]
\centering
\includegraphics[width=10cm]{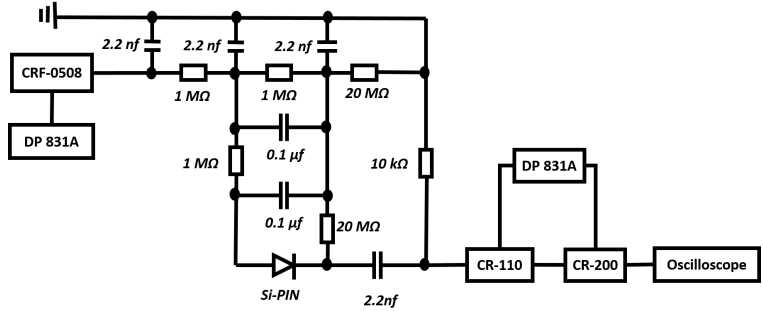}
\caption{\label{readout} The circuit diagram of the readout system. DP 831A is a DC power supply made by Rigol company, CRF-0508 is a power module made by Tianjin Centre Advanced Technology Company, and CR-110 and CR-200 are pre-amplifier and shaping amplifier made by Cremat company. CRF-0508 converts an input voltage of 12~V to an output voltage of -700~V. }
\end{figure} 

(G)The radon detector consists of a 41.5~L stainless steel chamber, a pressure gauge (ISE80-A2-N-M, SMC corporation), a Si-PIN photodiode (S3204-09, Hamamatsu), and a set of electronic equipment~\cite{Previous_1, Previous_2, Previous_3}. The HV module and signal amplifiers are updated for remote operation and price consideration. The updated circuit diagram is shown in Fig.~\ref{readout}.  A power module (CRF-0508, Tianjin Centre Advanced Tech.~Co.~Ltd) is used to supply -700~V high voltage. A CR-110 pre-amplifier (Cremat. Inc.) and a CR-200 Gaussian shaping amplifier (Cremat. Inc.) are used to amplify and shape the signal. Two DC power suppliers (DP 831A, Rigol Technology Co.) are used to provide power to the power module and the amplifiers. DP 831A has supporting PC software to control the voltage loading, so the signal readout system can be operated remotely.

To get a lower background, the radon detector, the activated carbon container, and all the pipelines in the system are electro-polished to a roughness of 0.1~$\mu$m. Knife-edge flanges with metal gaskets and vacuum coupling radius seal (VCR) connectors with metal gaskets are used. The leakage rate of the system is better than 1$\times$10$^{-9}$~Pa$\cdot$m$^3$/s, which is measured by a helium leak detector (ZQJ-3000, KYKY Technology Co. Ltd).

\section{Adsorption coefficient measurement} \label{sec3}

\begin{figure}[htbp]
\centering
\includegraphics[width=10cm]{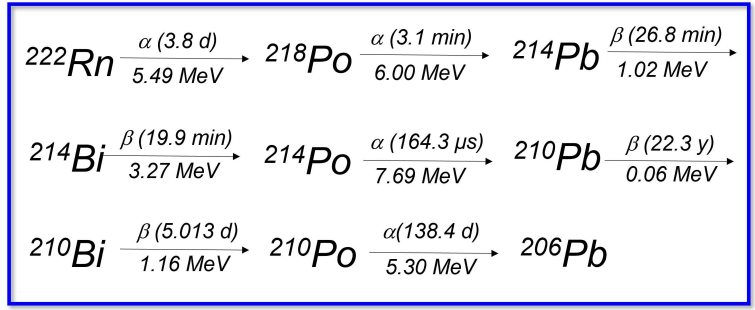}
\caption{\label{decaychain} The relevant branches of $^{222}$Rn decay chain. The half-lives of the nuclide are in parentheses. The $\alpha$ decay energy or the maximum $\beta$ decay energy are indicated below the arrow. }
\end{figure} 

\begin{figure}[htbp]
\centering
\includegraphics[width=6.5cm]{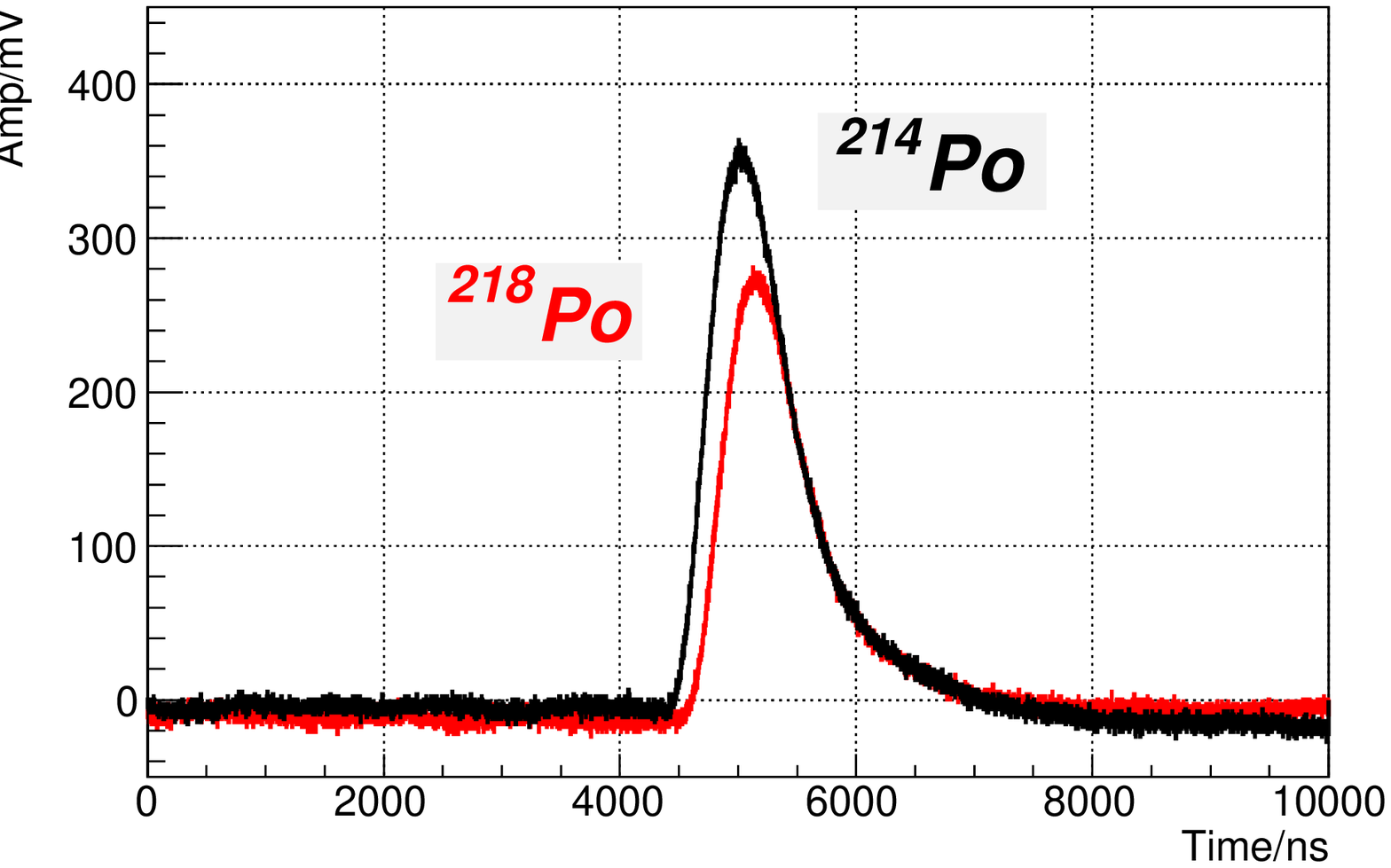}
\includegraphics[width=6.5cm]{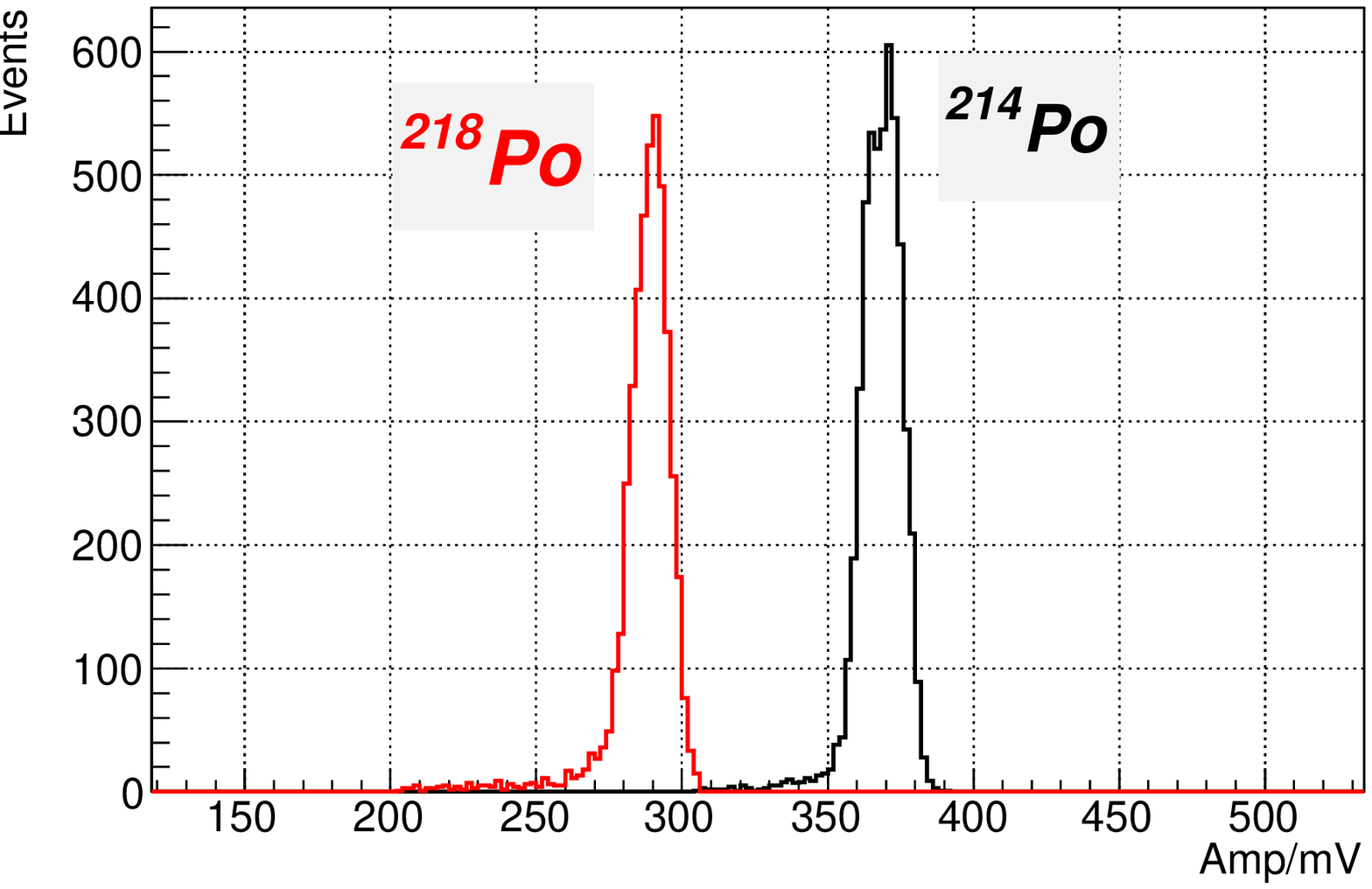}
\caption{\label{222Rn}Left: Examples of signal pulse. Right: An energy spectrum of $^{222}$Rn source. }
\end{figure}

Activated carbon is widely used in low-background experiments for radon removal. The radon adsorption coefficient of activated carbon is used to quantify its radon adsorption capability. Fig.~\ref{decaychain} shows the relevant branches of $^{222}$Rn decay chain and the radon concentration is determined by detecting the $\alpha$s from $^{218}$Po and $^{214}$Po decay. The left of Fig.~\ref{222Rn} shows example pulses of $^{218}$Po and $^{214}$Po and the right of Fig.~\ref{222Rn} shows an energy spectrum of a $^{222}$Rn source. With the $^{222}$Rn source calibration data, the calibration factor, which represents the detecting efficiency, can be derived. The calibration factor for $^{218}$Po is 60.4 $\pm$6.0~cph/(Bq/m$^3$) and is 67.0 $\pm$6.7~cph/(Bq/m$^3$) for $^{214}$Po. Because $^{214}$Po is the decay substrate produced by $^{218}$Po via $^{214}$Pb and $^{214}$Bi, the radon detector has a higher detection efficiency for $^{214}$Po.  The detail of the detector and the detecting principle can be found in Ref.~\cite{Previous_1, Previous_2, Previous_3}. 

The adsorption coefficient is calculated with the $^{222}$Rn breakthrough curve, which is measured as follows:

(A) Turn on the temperature control system to heat the activated carbon to 200$^{\circ}$C and purge the system with evaporative nitrogen for 2 hours. The radon concentration in evaporated nitrogen is of the order of 0.1~mBq/m$^3$, and the activated carbon does not have adsorption capacity for radon gas at 200$^{\circ}$C, so this step can effectively remove the radon gas adsorbed by the activated carbon during the storage and reduce the measurement background.

(B) Stop purging and vacuum the system to -101~kPa. This process can eliminate the effect of evaporated nitrogen on the measurement.

(C) Cool the activated carbon to a specified temperature with the temperature control system. The adsorption capacity of activated carbon for radon is temperature dependent, so the activated carbon must be controlled to a specified temperature before the measurement.

(D) Open the relevant valves and use the MFCs to control the gas passing through the activated carbon to a specified flow rate. The difference between the flow rates set for MFC1 and MFC2 is the gas flow rate through the activated carbon.

(E) Open the outlet valve of the radon detector after its relative internal pressure is larger than 10~kPa. The purpose of opening the valve when the air pressure is higher than atmospheric pressure is to prevent the radon gas in the air from diffusing into the detector, thus affecting the measurement results.

(F)Monitor the radon concentration inside the detector until it reaches equilibrium. At the beginning of the experiment, all the radon in the gas will be adsorbed by the activated carbon, but as the experiment proceeds, the activated carbon is gradually saturated, so its adsorption efficiency will gradually decrease. When the radon concentration in the detector reaches equilibrium, it means the radon adsorption efficiency of activated carbon drops to zero.

As is shown in Fig.~\ref{decaychain}, the half-life of $^{218}$Po is only 3.1~min, it can reach dynamic equilibrium quickly, so the $^{218}$Po counting rate is used to characterize the $^{222}$Rn concentration inside the detector. 

\begin{figure}[htbp]
\centering
\includegraphics[width=8cm]{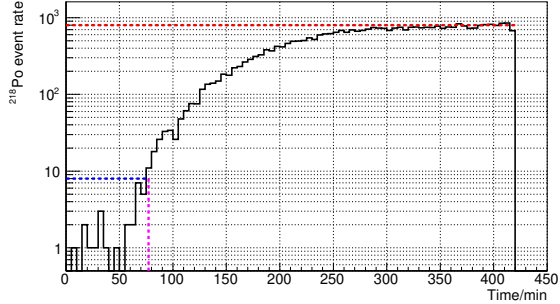}
\caption{\label{adsorptionexample}  $^{222}$Rn breakthrough curve. The Y-axis is the event rate of $^{218}$Po. This curve is measured at -80$^{\circ}$C with 0.73~g of activated carbon, and the gas flow rate is 4~L/min. The average event rate at equilibrium is indicated with the red dashed line, and the 1$\%$ event rate is indicated with the blue dashed line. The intersection of the pink dashed line and the X-axis is when the radon adsorption efficiency decreases to 99\%, which is used to calculate the radon adsorption coefficient. The events in the first 50 minutes are caused by the residual $^{222}$Rn from the last test.}
\end{figure}

In this work, 0.73~g of activated carbon is used. Fig.~\ref{adsorptionexample} is an example of the $^{222}$Rn breakthrough curve measured at -80~$^\circ$C. The radon concentration in the gas is 160 $\pm$ 16~Bq/m$^3$, which is derived from the $^{218}$Po counting rate and the $^{218}$Po calibration factor. The gas flow rate for this test is 4~L/min.

In the first 55 minutes, we did not observe an event rate increase, which means basically the radon adsorption efficiency of activated carbon is $\sim$100\%. From 55 to 350 minutes, the event rate gradually increases, which means the radon adsorption efficiency of activated carbon gradually decreases, and the efficiency drops to zero at 350 minutes. The events in the first 50 minutes are caused by the residual radon of the last measurement. 

The average event rate at equilibrium is indicated with a red dashed line, which is 805 counts/5min, and the 1$\%$ event rate is indicated with a blue dashed line, which is 8 events/5min. 1\% event rate means 99\% radon adsorption efficiency. Thus, the intersection of the pink dashed line and the X-axis is when the radon adsorption efficiency of the activated carbon decreases to 99\%. The time is used to calculate the radon adsorption coefficient, which is defined according to:

\begin{equation}
C = \frac{F_G \times t_A}{m_{AC}}
\label{Eq.1}
\end{equation}

Where C is the radon adsorption coefficient in the unit of L/g, F$_G$ is the gas flow rate in the unit of L/min, and t$_A$ is the time when the event rate increases to 1\% of the equilibrium event rate which means the radon adsorption efficiency decrease to 99\%, and its unit is minute, m$_{AC}$ is the activated carbon mass in the unit of gram which is 0.73~g in this measurement.

According to Fig.~\ref{adsorptionexample}, the radon adsorption coefficient at -80 $\pm$ 5~$^\circ$C is 425 $\pm$ 14 ~L/g, and the error is derived from the gas flow rate and the bin width of the histogram.

\section{Influencing factors of radon adsorption on activated carbon } \label{sec4}
The intrinsic background and its radon adsorption capability are the main factors considered in the application of activated carbon. According to our previous work~\cite{Previous_1}, the $^{226}$Ra content of the SARATECH polymer-based spherical activated carbon is $\sim$1.36~mBq/kg, which can be used for radon enrichment. While the adsorption capability of the activated carbon is not only affected by its characteristics, micro-porous structure, and relative surface, for example, but also by external conditions. In this work, the authors mainly consider the temperature, the gas flow rate, and the radon concentration in the gas.

\subsection{Influencing factor of temperature} \label{sec4.1}
The adsorption capability of activated carbon on radon at low temperatures will be significantly enhanced. However, according to our limited knowledge, there is no experiment to give the optimal temperature for radon adsorption on activated carbon.

\begin{figure}[htbp]
\centering
\includegraphics[width=8cm]{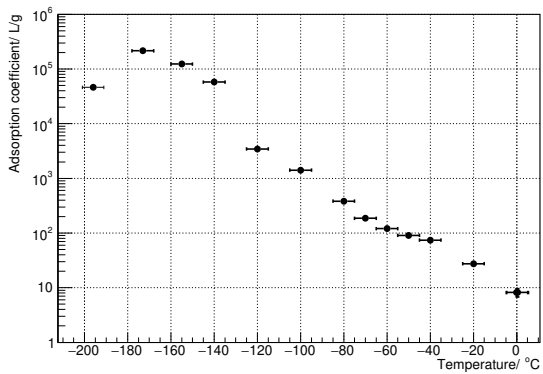}
\caption{\label{T_charcoal}  The dependence of adsorption coefficient on temperature. During the test, the radon concentration in the gas is 160 $\pm$ 16~Bq/m$^3$. For the tests between 0~$^\circ$C and -60~$^\circ$C, the gas flow rate is ~1~L/min; for the tests between -70~$^\circ$C to -120~$^\circ$C, the gas flow rate is 4~L/min, for other tests, the gas flow rate is 15~L/min.
 }
\end{figure} 

In this work, the adsorption coefficient dependence on temperature has been measured and the measurement results are shown in Fig.~\ref{T_charcoal}. During the measurements, the radon concentration in the gas is 160 $\pm$ 16~Bq/m$^3$. For the tests between 0~$^\circ$C and -60~$^\circ$C, the gas flow rate is ~1~L/min; for the tests between -70~$^\circ$C to -120~$^\circ$C, the gas flow rate is 4~L/min, for others, the gas flow rate is 15~L/min.

As can be seen from Fig.~\ref{T_charcoal}, from -173~$^\circ$C to 0 ~$^\circ$C, the radon adsorption capability of activated carbon increased significantly along with the decrease of temperature. However, when the activated carbon is cooled to -196~$^\circ$C, the adsorption capability sharply decreases. This is because when the temperature drops to -196~$^\circ$C, the carrier gas, nitrogen, is liquefied, and the adsorption capability of activated carbon to radon gas in a liquid is significantly lower than that in gas. Therefore, we can conclude that for the radon adsorption of activated carbon, the lower the temperature is, the stronger the adsorption capability is. Still, it must not be lower than the liquefaction point of the carrier gas.

\subsection{Influencing factor of gas flow rate}

During the measurement described in Sec.~\ref{sec4.1}, three different values of gas flow rate are used. When the temperature is relatively high, the radon adsorption capability of activated carbon is weak, and using a large gas flow rate at this time will introduce a large error. While the temperature is low, the radon adsorption capability is strong, a small gas flow rate will make the test time too long.

\begin{figure}[htbp]
\centering
\includegraphics[width=8cm]{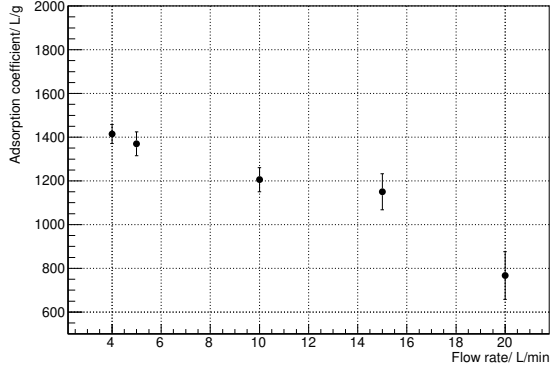}
\caption{\label{flowrate}  The dependence of adsorption coefficient on gas flow rate. During the test, the radon concentration in the gas is 160 $\pm$ 16~Bq/m$^3$, and the temperature is -100 $\pm$ 5~$^\circ$C.  
 }
\end{figure}

Five different gas flow rates with the same $^{222}$Rn concentration at -100 $\pm$ 5~$^\circ$C have been used to study the influence of gas flow rate on the adsorption coefficient. As is shown in Fig.~\ref{flowrate}, the adsorption coefficient decreases along with the increase of the gas flow rate. This may be caused by insufficient contact between radon and activated carbon at a high flow rate. 

\subsection{Influencing factor of radon concentration}

\begin{figure}[htbp]
\centering
\includegraphics[width=8cm]{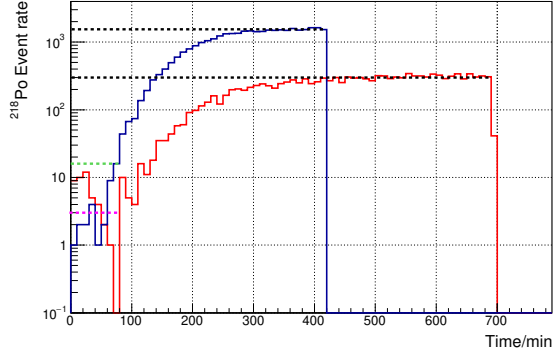}
\caption{\label{DC}  Two breakthrough curves for different radon concentrations. During the tests, the temperature is -80 $\pm$ 5$^\circ$C, and the flow rate is 4~L/min. While for the $^{222}$Rn concentration, the blue curve is 160 $\pm$ 16~Bq/m$^3$ and the red curve is 32.0 $\pm$ 3.2~Bq/m$^3$. The two black dotted lines indicate the $^{218}$Po event rate at equilibrium for two conditions, and the green and pink dotted lines indicate the 1\% event rate for two conditions, respectively. The events before 70~minutes are caused by the residual $^{222}$Rn from the previous measurement.
 }
\end{figure}

Except for the temperature and gas flow rate, the dependence of the adsorption coefficient on $^{222}$Rn concentration has also been measured. Two different radon concentrations have been used in the tests, and the results are shown in Fig.~\ref{DC}. The blue curve represents the $^{222}$Rn concentration in the gas is of 160 $\pm$ 16~Bq/m$^3$, and the red curve represents the $^{222}$Rn concentration in the gas is of 32.0 $\pm$ 3.2~Bq/m$^3$. The two black dotted lines indicate the $^{218}$Po event rate at equilibrium for two conditions, and the green and pink dotted lines indicate the 1\% event rate for two conditions, respectively. The events before 70~minutes are caused by the residual $^{222}$Rn from the previous measurement. During the measurements, the temperature is -80 $\pm$ 5$^\circ$C, and the flow rate is 4~L/min.

According to Fig.~\ref{DC}, under these two different $^{222}$Rn concentrations, the breakthrough curves both reach equilibrium at $\sim$400 minutes and get the 1\% event rate at $\sim$80 minutes, namely, no clear correlation between the adsorption coefficient and $^{222}$Rn concentration has been observed.

\section{Sensitivity estimation}\label{sec5}

The $^{222}$Rn concentration measurement sensitivity of this system depends on the sensitivity of the $^{222}$Rn detector and the enrichment ability of the low-temperature activated carbon system. The sensitivity of the $^{222}$Rn detector is 0.71~mBq/m$^3$ for a one-day measurement and the details can be found in Ref.~\cite{Previous_1}.



The enrichment ability of the activated carbon system depends on the adsorption coefficient and the $^{222}$Rn detector volume. The desorption efficiency of activated carbon at 200~$^\circ$C is generally considered to be $\sim$100\%~\cite{Borexino}. The enrichment factor can be obtained by dividing the volume of gas flow through the activated carbon by the volume of the $^{222}$Rn detector. The gas volume purged through the activated carbon should not saturate it with $^{222}$Rn. The adsorption coefficient has to be taken into account. The system shown in Fig.~\ref{detector} is used here. The gas flow rate is set to 10~L/min, the $^{222}$Rn concentration is $\sim$2.5~Bq/m$^3$, 0.73~g of activated is used, and the temperature is -140 $\pm$ 5$^\circ$C. The volume of $^{222}$Rn containing gas purged through activated carbon is 10~m$^3$, which does not saturate the activated carbon according to the adsorption coefficient shown in Fig.~\ref{T_charcoal}. The operating procedures are as below:

(A)-(E) Same as the adsorption coefficient measurement discussed in Sec.~\ref{sec3}.

(F) Monitor the gas volume flow through the activated carbon and close valve1 and valve2 after the gas volume is 10~m$^3$. The gas volume is determined by the expected enrichment factor but cannot exceed the maximum adsorption volume of the activated carbon.

(G) Pump the $^{222}$Rn detector to -101~kPa. This step ensures that all the adsorbed gas is transferred into the detector and also allows the detector to operate at atmospheric pressure. 

(H) Turn on the temperature control system to heat the activated carbon to 200~$^\circ$C and open valve1 and valve2 after the temperature is stabilized to 200~$^\circ$C. The purpose of opening the valve after the temperature has stabilized at 200~$^\circ$C is to ensure that all the adsorbed radon can be desorbed.

(I) Seal the $^{222}$ Rn detector after the pressure reaches 0~kPa. The $^{222}$Rn detection efficiency is related to the internal pressure of the detector. The detector in this work is calibrated at atmospheric pressure, so the internal pressure needs to be controlled at $\sim$0~kPa.

(J) Measure the $^{222}$Rn concentration in the detector.

\begin{figure}[htbp]
\centering
\includegraphics[width=8cm]{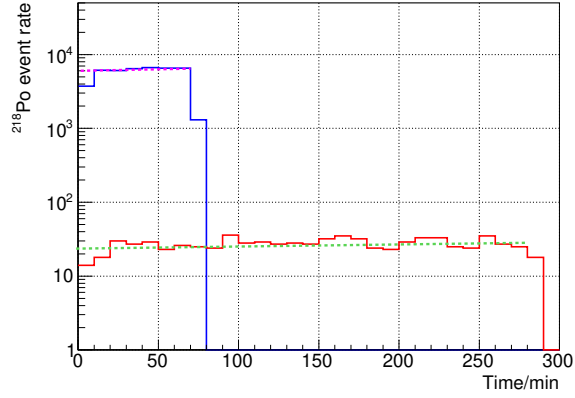}
\caption{\label{CE} The $^{218}$Po event rate before and after the enrichment. The red curve is the direct measurement result, and the average event rate, indicated with a green dashed line, is 28.3 $\pm$ 1.0 counts/10min. The blue curve is the event rate after enrichment, and the average event rate, indicated with a pink dashed line, is 6444 $\pm$ 36 counts/10min.
 }
\end{figure} 

 Fig.~\ref{CE} is the $^{218}$Po event rates before and after enrichment. The red curve is the direct measurement result of $^{222}$Rn concentration in gas, and the blue curve is after enrichment. The expected $^{218}$Po event rate can be calculated according to:

\begin{equation}
R_E = R_0 \times (\frac{V_G}{V_D} + 1) \times C_D  
\label{Eq.3}
\end{equation}

Where R$_E$ is the expected event rate after enrichment, R$_0$ is the direct measurement event rate, V$_G$ is the gas volume flow through the activated carbon, V$_D$ is the detector volume, and one is because one volume of gas will be used to bring $^{222}$Rn from the activated carbon chamber into the detector, C$_D$ is the decay factor, which is calculated with:

\begin{equation}
C_D = \frac{\int_{0}^{t_E}e^{-{\lambda}t}dt}{t_E}
\label{Eq.4}
\end{equation}

Where $\lambda$ is the $^{222}$Rn decay constant, t$_E$ is the enrich time. In this measurement, t$_E$ = 1000 min, C$_D$ = 0.939, R$_0$ = 28.3 $\pm$ 1.0 counts/10min. R$_0$ is calculated with the event rate from 20~min to 270~min shown in Fig.~\ref{CE}. Thus R$_E$ is expected to be (6.43 $\pm$ 0.24) $\times$ 10$^3$ counts/10min. The measurement result is in good accordance with R$_E$, which is (6.444 $\pm$ 0.036) $\times$ 10$^3$ counts/10min. 

The $^{222}$Rn concentration measurement sensitivity can be estimated with Eq.~\ref{Eq.5}:

\begin{equation}
L = \frac{L_D}{(\frac{V_G}{V_D} + 1) \times C_D}
\label{Eq.5}
\end{equation}

Where L is the sensitivity of the system, L$_D$ is the sensitivity of the $^{222}$Rn detector which is 0.71~mBq/m$^3$, V$_G$, V$_D$, and C$_D$ are the same as in Eq.~\ref{Eq.3}. With the setup described above, the sensitivity of the system is $\sim$3$~\mu$Bq/m$^3$.

\section{Discussion}\label{sec6}
\subsection{Adsorption coefficient}
According to the measurement results described in Sec.~\ref{sec4}, the adsorption coefficient of activated carbon depends on the temperature and gas flow rate. In the measurement shown in Fig.~\ref{T_charcoal}, the authors used different gas flow rates at different temperatures in order to optimize measurement accuracy and reduce test duration. Therefore, the adsorption coefficient can be further increased if a small gas flow rate is used.

\subsection{Activated carbon background}
According to our measurement, the $^{226}$Ra concentration in Saratech activated carbon is $\sim$1.36~mBq/kg. The activity of $^{222}$Rn emanated from the activated carbon can be calculated with Eq.~\ref{Eq.5}:

\begin{equation}
A_{Rn} = A_{Ra} \times (1-e^{-\lambda t_E})
\label{Eq.5}
\end{equation}

Where A$_{Rn}$ is the $^{222}$Rn activity, A$_{Ra}$ is the $^{226}$Ra activity of the activated carbon, $\lambda$ is the $^{222}$Rn decay constant, t$_E$ is the enrich time.

If t$_E$=24~h, the background contributed by the activated carbon will be $\sim$0.23~$\mu$Bq. Therefore, the influence of the activated carbon background on the measurement is negligible. 

\subsection{Limit sensitivity estimation}

As described in Sec.~\ref{sec5}, if the temperature is $\sim$-140~$^\circ$C, the gas flow rate is 10~L/min, and the enriching time is 1000~min, the sensitivity is $~\sim$3$~\mu$Bq/m$^3$. While according to Fig.~\ref{T_charcoal}, 1~g of activated carbon can be used to adsorb the radon in more than 200~m$^3$ radon gas. With the parameters shown in Tab.~\ref{tab1}, the limit sensitivity of the system can be calculated according to Eq.~\ref{Eq.4}, and it is $\sim$0.3~$\mu$Bq/m$^3$.

When T$_E$ = 9.26~d, the $^{222}$Rn background contributed by the activated is $\sim$1.1~$\mu$Bq, if averaged to the 200~m$^3$ gas, it will be 0.0055~$\mu$Bq/m$^3$, which is negligible to the measurement.

\begin{table}[htb]
\centering
\begin{tabular}{ccccccc}
	\hline
	Temperature( $^\circ$C) & F$_G$(L/min) & T$_E$(min) & V$_G$(m$^3$) & C$_D$ & \\ \hline
	-173 $\pm$ 5 & 15 & 9.26 $\times$ 24 $\times$ 60  & 200 & 0.48 & \\ \hline
\end{tabular}
\caption{\label{tab1} Parameters used for limit sensitivity estimation. 
}
\end{table}

\section{Conclusions and future prospects}
JUNO is a multipurpose neutrino experiment aiming to determine the neutrino mass ordering and precisely measure the neutrino oscillation parameters. $^{222}$Rn is one of the main background sources. To measure the $^{222}$Rn concentration in the nitrogen used for JUNO, a highly sensitive $^{222}$Rn concentration measurement system based on $^{222}$Rn enrichment of low-temperature activated carbon and $^{222}$Rn measurement of the electrostatic collection has been developed and the limit sensitivity of the system is $\sim$0.3~$\mu$Bq/m$^3$. 

In this paper, the adsorption coefficient of Saratech activated carbon is also measured at different conditions. The results show that above the liquefaction temperature of the carrier gas, the lower the temperature, the larger is the adsorption coefficient. Besides, the adsorption coefficient also increases with the decrease of the gas flow rate.

The measurement results also show that the Saratech activated carbon is an excellent candidate for the research and development of radon removal devices. The amount of activated carbon, working temperature, and gas flow rate need to be determined according to specific experimental conditions.

\section{Acknowledgments}

This work is supported by the National Natural Science Foundation of China - Yalong River Hydropower Development Co., LTD. Yalong River Joint Fund (Grant No. U1865208), the Strategic Priority Research Program of the Chinese Academy of Sciences (Grant No. XDA10011200), the Innovative Project of the Institute of High Energy Physics (Grant No. Y954514), and the National Natural Science Foundation of China (Grant No. 11875280, No. 11905241).



\begin{thebibliography}{00}
\bibitem{JUNO_1} 
JUNO Collaboration, Neutrino physics with JUNO, J. Phys. G: Nucl. Part. Phys. 43 (2016) 030401.
\bibitem{JUNO_2} 
JUNO Collaboration, JUNO physics and detector, Progress in Particle and Nuclear Physics 123 (2022) 103927.
\bibitem{LS}
P.~Lombardi, M.~Montuschi, A.~Formozov, et al., Distillation and stripping pilot plants for the JUNO neutrino detector: Design, operations and reliability, Nucl. Instrum. Meth. A 925 (2019) 6-17.
\bibitem{JUNO_water}
C.~Guo, J.C.~Liu, Y.P.~Zhang, P.~Zhang, C.G.~Yang, Y.B.~Huang, W.X.~Xiong, H.Q.~Zhang, Y.T.~Wei, Y.Y.~Gan, Study on the radon removal for the water system of Jiangmen Underground Neutrino Observatory, RDTM (2018)2:48.
\bibitem{Previous_1}
Y.Y.~Chen, Y.P.~Zhang, Y.~Liu, J.C.~Liu, C.~Guo, P.~Zhang, S.K.~Qiu, C.G.~Yang, and Q.~Tang, A study on the radon removal performance of low background activated carbon, JINST 17 (2022) P02003.
\bibitem{AC_1}
M.~Wojcik, G.~Zuzel, Low-$^{222}$Rn nitrogen gas generator for ultra-low background counting systems,  Nucl. Instrum. Meth. A 539 (2005) 427-432.
\bibitem{AC_2}
K.~Pushkin, C.~Akerlof, D.~Anbajagane, J.~Armstrong, M.~Arthurs, J.~Bringewatt, T.~Edberg, C.~Hall, M.~Lei, R.~Raymond, M.~Reh, D.~Saini, A.~Sander, J.~Schaefer, D.~Seymour, N.~Swanson, Y.~Wang, W.~Lorenzon, Study of radon reduction in gases for rare event search experiments,  Nucl. Instrum. Meth. A 903 (2018) 267-276.
\bibitem{SuperK_1}
Y.~Takeuchi, K.~Okumura, T.~Kajita, S.~Tasaka, H.~Hori, M.~Nemoto, et al., Development of high sensitivity radon detectors, Nucl. Instrum. Meth. A 421 (1999) 334.
\bibitem{SuperK_2}
Y.~Nakano, H.~Sekiya, S.~Tasaka, Y.~Takeuchi, R.A.~Wendell, M.~Matsubara, M.~Nakahata, Measurement of radon concentration in super-Kamiokande's buffer gas, Nucl. Instrum. Meth. A 867 (2017) 108-114.
\bibitem{SuperK_3}
Y.~Nakano a, T.~Hokama, M.~Matsubara, M.~Miwa, M.~Nakahata, T.~Nakamura, H.~Sekiya, Y.~Takeuchi, S.~Tasaka, R.A.~Wendell, Measurement of the radon concentration in purified water in the Super-Kamiokande IV detector,  Nucl. Instrum. Meth. A 977 (2020) 164297.
\bibitem{Borexino_2}
J.~Kiko, Detector for $^{222}$Rn measurements in air at the 1mBq/m$^3$ level, Nucl. Instrum. Meth. A 460 (2001) 272-277.
\bibitem{DS}
DarkSide collaboration, DarkSide-50 532-day Dark Matter Search with Low-Radioactivity Argon, Phys. Rev. D 98 (2018) 102006.
\bibitem{SuperK}
Super-Kamiokande collaboration, The Super-Kamiokande detector, Nucl. Instrum. Meth. A 501 (2003)418.
\bibitem{Xmass}
XMASS collaboration, Radon removal from gaseous xenon with activated charcoal, Nucl. Instrum. Meth. A 661 (2012) 50.
\bibitem{ACfiber}
Y.~Nakano, K.~Ichimura, H.~Ito, T.~Okada, H.~Sekiya, Y.~Takeuchi, S.~Tasaka, and M.~Yamashita, Evaluation of radon adsorption efficiency values in xenon with activated carbon fibers, Prog. Theor. Exp. Phys. 2020, 113H01.
\bibitem{Borexino}
G.~Heusser, W.~Rau, B.~Freudiger, M.~Laubenstein, M.~Balata, T.~Kirsten, $^{222}$Rn detection at the $\mu$Bq/m$^3$ range in nitrogen gas and a new Rn purification technique for liquid nitrogen, Applied Radiation and Isotopes 52 (2000) 691-695.
\bibitem{Peking}
Lu Guo, Yunxiang Wang, Lei Zhang, Zhi Zeng, Wenbin Dong, Qiuju Guo, The temperature dependence of adsorption coefficients of $^{222}$Rn on activated charcoal: an experimental study, Applied Radiation and Isotopes 125 (2017) 185-187.
\bibitem{Saratech}
Saratech charcoal specification, https://www.bluecher.com/en/.
\bibitem{Previous_2}
Y.P.~Zhang, J.C.~Liu, C.~Guo, Y.B.~Huang, C.~Xu, M.Y.~Guan, C.G.~Yang, P.~Zhang, The development of $^{222}$Rn detectors for JUNO prototype, RDTM (2018) 2:5.
\bibitem{Previous_3}
L.F.~Xie, J.C.~Liu, S.K.~Qiu, C.~Guo, C.G.~Yang, Q.~Tang, Y.P.~Zhang, P.~Zhang, Developing the radium measurement system for the water Cherenkov detector of the Jiangmen Underground Neutrino Observatory,  Nucl. Instrum. Meth. A 976 (2020) 164266.









\end{thebibliography}



\end{document}